\documentclass[letterpaper, 10 pt, conference]{ieeeconf}  

\IEEEoverridecommandlockouts                              
\usepackage{graphics} 
\usepackage{epsfig} 
\usepackage{mathptmx} 
\usepackage{times} 
\usepackage{amsmath} 
\usepackage{amssymb}  
\usepackage{floatrow}
\usepackage{multirow}
\usepackage{algorithm}
\usepackage{algpseudocode}
\usepackage{color}

\newtheorem{remark}{\indent Remark}

\title{\LARGE \bf
Sampling-based Learning Control for Quantum Systems with Hamiltonian Uncertainties}
\author{Daoyi~Dong, Chunlin~Chen, Ruixing~Long, Bo Qi, Ian R.~Petersen
\thanks{This work was supported by the Australian Research Council (DP130101658, FL110100020)
and by the Natural Science
Foundation of China under Grant No.61273327.}
\thanks{D. Dong  and I. R. Petersen are with the School of Information Technology and Electrical
Engineering, University of New South Wales at the Australian Defence
Force Academy, Canberra, ACT 2600, Australia {\tt\small daoyidong@gmail.com}; {\tt\small i.r.petersen@gmail.com}.}
\thanks{C. Chen is with the Department of Control and System Engineering, Nanjing
University, Nanjing 210093, China and with the Department of
Chemistry, Princeton University, Princeton, New Jersey 08544, USA
{\tt\small clchen@nju.edu.cn}.}
\thanks{R. Long is with the Department of
Chemistry, Princeton University, Princeton, New Jersey 08544, USA
{\tt\small ruixing.long@gmail.com}.}
\thanks{B. Qi is with the Key Laboratory of Systems and Control,
Academy of Mathematics and Systems Science, Chinese Academy
of Sciences, Beijing 100190, China {\tt\small qibo@amss.ac.cn}.}
}

\begin{document}

\maketitle
\begin{abstract}
Robust control design for quantum systems has been recognized as a key task in the development of practical quantum technology. In this
paper, we present a systematic numerical methodology of sampling-based
learning control (SLC) for control design of quantum systems with Hamiltonian uncertainties.
The SLC method includes two steps of ``training" and ``testing and
evaluation". In the training step, an augmented system is
constructed by sampling uncertainties according to possible
distributions of uncertainty parameters. A gradient flow based
learning and optimization algorithm is adopted to find the control
for the augmented system. In the process of testing and
evaluation, a number of samples obtained through sampling the uncertainties are tested to
evaluate the control performance. Numerical results
demonstrate the success of the SLC
approach. The SLC method has potential applications for robust control design of
quantum systems.
\end{abstract}


\section{Introduction}\label{Sec1}
Controlling quantum phenomena lies at the heart of quantum technology and quantum
control theory is drawing wide interests from scientists and engineers
\cite{Dong and Petersen 2010IET}-\cite{Brif et al 2010}. In recent years, robust control of quantum systems has been recognized as a key requirement
for practical quantum technology since the existence of uncertainties is unavoidable in the modeling and control process for real quantum
systems \cite{Pravia et al 2003}-\cite{James 2004}. Several methods have been
proposed for robust control design of quantum systems. For example,
James \emph{et al}. \cite{James et al 2007} have formulated and
solved an $H^{\infty}$ controller synthesis problem for a class of
quantum linear stochastic systems in the Heisenberg picture. Dong
and Petersen \cite{Dong and Petersen 2009NJP}-\cite{Dong and Petersen 2011IFAC} have proposed a sliding mode control
approach to deal with Hamiltonian uncertainties in two-level quantum systems. Chen
\emph{et al}. \cite{Chen et al 2012} have proposed a fuzzy estimator based approach
for robust control design in quantum systems.

In this paper, we present a systematic numerical methodology for control
design of quantum systems with Hamiltonian uncertainties.
The proposed method includes two steps: ``training" and ``testing
and evaluation", and we call it sampling-based learning control
(SLC). In the training step, we sample the uncertainties according to
possible distributions of uncertainty parameters and construct an
augmented system using these samples. Then we develop a gradient
flow based learning and optimization algorithm to find the control
with desired performance for the augmented system. In the
process of testing and evaluation, we test a number of
samples of the uncertainties to evaluate the control performance. Numerical
results show that the SLC method is useful for
control design of quantum systems with Hamiltonian uncertainties.

This paper is organized as follows. Section \ref{Sec2} formulates
the control problem. Section \ref{Sec3}  presents the
approach of sampling-based learning control and introduces  a gradient
flow based learning and optimization algorithm. A result on
control design in three-level quantum systems is
presented in Section \ref{Sec4}. Concluding remarks are
presented in Section \ref{Sec5}.

\section{Model and problem formulation}\label{Sec2}
We focus on finite-dimensional closed quantum systems. For a finite-dimensional closed quantum system, the
evolution of its state $|\psi(t)\rangle$ can be described by the
following Schr\"{o}dinger equation (setting $\hbar=1$):
\begin{equation} \label{systemmodel}
\left\{ \begin{array}{l}
  \frac{d}{dt}|{\psi}(t)\rangle=-iH(t)|\psi(t)\rangle \\
 t\in [0, T], \ |\psi(0)\rangle=|\psi_{0}\rangle.\\
\end{array}
\right.
\end{equation}
The dynamics of the system are governed by a
time-dependent Hamiltonian of the form
\begin{equation}\label{Hamiltonian}
H(t)=H_{0}+H_{c}(t)=H_{0}+\sum_{m=1}^{M}u_{m}(t)H_{m},
\end{equation}
where $H_{0}$ is the free Hamiltonian of the system,
$H_{c}(t)=\sum_{m=1}^{M}u_{m}(t)H_{m}$ is the time-dependent control
Hamiltonian that represents the interaction of the system with the
external fields $u_{m}(t)$, and the $H_{m}$ are Hermitian operators through which the
controls couple to the system.

The solution of (\ref{systemmodel}) is given
by $\displaystyle |\psi(t)\rangle=U(t)|\psi_{0}\rangle$, where the
propagator $U(t)$ satisfies
\begin{equation}
\left\{ \begin{array}{c}
  \frac{d}{dt}U(t)=-iH(t)U(t),\\
  t\in [0, T], \ U(0)=\textrm{Id}.\\
\end{array}
\right.
\end{equation}

For an ideal model, there exist no uncertainties in (\ref{Hamiltonian}). However, for a practical quantum system, the existence of uncertainties is unavoidable. In this paper, we consider that the system Hamiltonian has the following form
\begin{equation}
H_{\omega, \theta}(t)=g(\omega(t))H_{0}+\sum_{m=1}^{M}f(\theta(t))u_{m}(t)H_{m}.
\end{equation}
The functions $g(\omega(t))$ and $f(\theta(t))$ characterize possible Hamiltonian uncertainties. We assume that the parameters
$\omega(t)$ and $\theta(t)$  are time-dependent, $\omega(t)\in [-\Omega, \Omega]$ and $\theta(t)\in [-\Theta, \Theta]$. The constants
$\Omega \in [0,1]$ and $\Theta \in [0,1]$ represent the bounds of
the uncertainty parameters. Now the objective is to design the controls
$\{u_{m}(t), m=1,2,\ldots , M\}$ to steer the
quantum system with Hamiltonian uncertainties from an initial state $|\psi_{0}\rangle$ to a target
state $|\psi_{\text{target}}\rangle$ with high fidelity. The control
performance is described by a \emph{performance function} $J(u)$ for
each control strategy $u=\{u_{m}(t), m=1,2,\ldots , M\}$. The
control problem can then be formulated as a maximization problem as
follows:

\begin{equation}\label{ensemble control}
\begin{split}
\displaystyle \max_u \ \ \  & J(u):=\vert
\langle\psi(T)|\psi_{\text{target}}\rangle\vert^{2}\\
\text{s.t.} \ \ \ & \frac{d}{dt}|\psi(t)\rangle=-iH_{\omega,\theta}(t)|\psi(t)\rangle, \ |\psi(0)\rangle=|\psi_{0}\rangle \\
& H_{\omega,\theta}(t)=g(\omega(t))H_{0}+\sum_{m=1}^{M}f(\theta(t))u_{m}(t)H_{m},\\
& \textrm{ with } \omega(t) \in [-\Omega,\Omega], \ \ \theta(t) \in [-\Theta,\Theta],~ t \in [0, T].
\end{split}
\end{equation}
Note that $J(u)$ depends implicitly on the control
$u$ through the Schr\"odinger equation.

\section{Sampling-based learning control of quantum systems}\label{Sec3}
Gradient-based methods \cite{Brif et al 2010}, \cite{Long and Rabitz
2011}, \cite{Roslund and Rabitz 2009}
have been successfully applied to search for optimal solutions to a
variety of quantum control problems, including theoretical and
laboratory applications. In this paper, a gradient-based learning
method is employed to optimize the controls for quantum systems with uncertainties. However, it is impossible to directly calculate
the derivative of $J(u)$ since there exist Hamiltonian uncertainties. Hence
we present a systematic numerical methodology for control design
using some samples obtained through sampling the uncertainties.
These samples are artificial quantum systems whose Hamiltonians are determined
according to the distribution of the uncertainty parameters. Then the
designed control law is applied to additional samples to test and
evaluate the control performance. A similar idea has been used to design robust control pulses for electron shuttling \cite{Zhang et al 2012} and to design a control law for inhomogeneous quantum ensembles \cite{Chen et al 2013arXiv}. In this paper, a systematic sampling-based learning control method is
presented to design control laws for quantum systems with Hamiltonian uncertainties. This method includes two steps of
``training" and ``testing and evaluation".


\subsection{Sampling-based learning control}

\subsubsection{Training step}\label{sec:training}
In the training step, we obtain $N$ samples through sampling uncertainties according to the distribution (e.g., uniform
distribution) of the uncertainty parameters and then construct an
augmented system as follows
\begin{equation}\label{augmented-system}
\frac{d}{dt}\left(%
\begin{array}{c}
  |{\psi}_{\omega_1,\theta_1}(t)\rangle \\
  |{\psi}_{\omega_2,\theta_2}(t)\rangle \\
  \vdots \\
  |{\psi}_{\omega_N,\theta_N}(t)\rangle \\
\end{array}%
\right)
=-i\left(%
\begin{array}{c}
  H_{\omega_1,\theta_1}(t)|\psi_{\omega_1,\theta_1}(t)\rangle \\
  H_{\omega_2,\theta_2}(t)|\psi_{\omega_2,\theta_2}(t)\rangle \\
  \vdots \\
  H_{\omega_N,\theta_N}(t)|\psi_{\omega_N,\theta_N}(t)\rangle \\
\end{array}%
\right),
\end{equation}
where
$H_{\omega_n,\theta_n}=g(\omega_{n})H_{0}+\sum_{m}f(\theta_{n})u_{m}(t)H_{m}$
with $n=1,2,\dots,N$. The performance function for the augmented
system is defined by
\begin{equation}\label{eq:cost}
J_N(u):=\frac{1}{N}\sum_{n=1}^N J_{\omega_n,\theta_n}(u)=\frac{1}{N}\sum_{n=1}^{N}\vert \langle\psi_{\omega_n,\theta_n}(T)|\psi_{\text{target}}\rangle\vert^{2}.
\end{equation} The task in the training step is to find a control
strategy $u^*$
that maximizes the performance function defined in Eq.
\eqref{eq:cost}.

%

Assume that the performance function is $J_N(u^{0})$ with an initial
control strategy $u^{0}=\{u^{0}_{m}(t)\}$. We can apply the
gradient flow method to approximate an optimal control strategy
$u^{*}=\{u^{*}_{m}(t)\}$. The detailed gradient flow algorithm
will be presented in Subsection \ref{sec2.3}.

As for the issue of choosing $N$ samples,
we generally choose them according to possible distributions of the
uncertain parameters $\omega(t) \in [-\Omega,\Omega]$ and
$\theta(t) \in [-\Theta, \Theta]$. It is clear that the basic
motivation of the proposed sampling-based approach is to design the
control law using only a few samples instead of unknown uncertainties. Therefore, it is
necessary to choose the set of samples that are representative for
these uncertainties.

For example, if the distributions of both $\omega(t)$ and $\theta(t)$
are uniform, we may choose some equally spaced samples in the
$\omega-\theta$ space. For example, the intervals $[-\Omega,
\Omega]$ and $[-\Theta, \Theta]$ are divided into
$N_{\Omega}+1$ and $N_{\Theta}+1$ subintervals, respectively,
where $N_{\Omega}$ and $N_{\Theta}$ are usually positive odd
numbers. Then the number of samples $N=N_{\Omega}N_{\Theta}$,
where $\omega_{n}$ and $\theta_{n}$ can be chosen from the
combination of $(\omega_{n_{\Omega}}, \theta_{n_{\Theta}})$ as
follows

\begin{equation}\label{discrete}
\left\{ \begin{array}{c} \omega_{n} \in
\{\omega_{n_{\Omega}}=1-\Omega+\frac{(2n_{\Omega}-1)\Omega}{N_{\Omega}},
\ n_{\Omega}=1,2,\ldots, N_{\Omega}\},\\
\theta_{n} \in
\{\theta_{n_{\Theta}}=1-\Theta+\frac{(2n_{\Theta}-1)\Theta}{N_{\Theta}},\
\
n_{\Theta}=1,2,\ldots, N_{\Theta}\}. \\
\end{array}
\right.
\end{equation}
In practical applications, the numbers of $N_{\Omega}$ and
$N_{\Theta}$ can be chosen by experience or tried through
numerical computation. As long as the augmented system can model
the quantum system with uncertainties and is effective to find the optimal control
strategy, we prefer smaller numbers of $N_{\Omega}$ and
$N_{\Theta}$ to speed up the training process and simplify the
augmented system.

\subsubsection{Evaluation step}
In the step of testing and evaluation, we apply the optimized
control $u^{*}$ obtained in the training step to a large number of
samples through randomly sampling the uncertainties
and evaluate for each sample the control performance in terms of
the fidelity $F(|\psi(T)\rangle,|\psi_{\text{target}}\rangle)$ between the final state
$|\psi(T)\rangle$ and the target state
$|\psi_{\text{target}}\rangle$ defined as follows \cite{Nielsen
and Chuang 2000}
\begin{equation}\label{fidelity}
F(|\psi(T)\rangle,|\psi_{\text{target}}\rangle)=|\langle \psi(T)|\psi_{\text{target}}\rangle| .
\end{equation}
If the average fidelity for all the
tested samples are satisfactory, we accept the designed control
law and end the control design process. Otherwise, we should go
back to the training step and generate another optimized control
strategy (e.g., restarting the training step with a new initial
control strategy or a new set of samples).

\subsection{Gradient flow based learning and optimization algorithm}
\label{sec2.3}

To get an optimal control strategy $u^{*}=\{u^{*}_{m}(t), (t \in
[0,T]), m=1,2,\ldots, M\}$ for the augmented system
(\ref{augmented-system}), a good choice is to follow the
direction of the gradient of $J_N(u)$ as an ascent direction. For
ease of notation, we present the method for $M=1$. We introduce a
time-like variable $s$ to characterize different control
strategies $u^{(s)}(t)$. Then a gradient flow in the control space
can be defined as
\begin{equation}\label{gradientflowequation}
\frac{du^{(s)}}{ds} =\nabla J_N(u^{(s)}),
\end{equation}
where $\nabla J_N(u)$ denotes the gradient of $J_N$ with respect
to the control $u$. It is easy to show that if $u^{(s)}$ is the
solution of \eqref{gradientflowequation} starting from an
arbitrary initial condition $u^{(0)}$, then the value of $J_N$ is
increasing along $u^{(s)}$, i.e., $\frac{d}{ds}J_N(u^{(s)})\geq
0$. In other words, starting from an initial guess $u^{0}$, we
solve the following initial value problem
\begin{equation}\label{gradientflowequation2}
\left\{%
\begin{split}
  & \frac{du^{(s)}}{ds} = \nabla J_N(u^{(s)}) \\
  & u^{(0)}=u^{0} \\
\end{split}%
\right.
\end{equation}
in order to find a control strategy which maximizes $J_N$. This
initial value problem can be solved numerically by using a forward
Euler method over
the $s$-domain, i.e.,
\begin{equation}\label{iteration1}
u(s+\triangle s, t)=u(s,t)+\triangle s\nabla J_N(u^{(s)}).
\end{equation}

As for practical applications, we present its iterative
approximation version to find the optimal controls $u^*(t)$ in an
iterative learning way, where we use $k$ as an index of iterations
instead of the variable $s$ and denote the controls at iteration
step $k$ as $u^{k}(t)$.
Equation \eqref{iteration1} can be rewritten as
\begin{equation}\label{iteration2}
u^{k+1}(t)=u^{k}(t)+ \eta^{k}\nabla J_N(u^{k}),
\end{equation}
where $\eta^{k}$ is the updating step (learning rate in computer
science) for the $kth$ iteration. By \eqref{eq:cost}, we also obtain that
\begin{equation}
\nabla J_N(u)=\frac{1}{N}\sum_{n=1}^{N}\nabla J_{\omega_n,\theta_n}(u).
\end{equation}
Recall that $J_{\omega,\theta}(u)=\vert \langle\psi_{\omega,\theta}(T)\vert\psi_{\textrm{target}}\rangle\vert^2$ and $\vert\psi_{\omega,\theta}(\cdot)\rangle$ satisfies
\begin{equation}\label{app-eq:sch}
\frac{d}{dt}\vert\psi_{\omega,\theta}\rangle=-iH_{\omega,\theta}(t)\vert\psi_{\omega,\theta}\rangle,\quad \vert\psi_{\omega,\theta}(0)\rangle=\vert\psi_{0}\rangle.
\end{equation} For ease of notation, we consider the case where only one control is involved, i.e., $H_{\omega,\theta}(t)=g(\omega)H_0+u(t)f(\theta)H_1$. We now derive an expression for the gradient of $J_{\omega,\theta}(u)$ with respect to the control $u$ by using a first order perturbation. Let $\delta\psi(t)$ be the modification of $\vert \psi(t)\rangle$ induced by a perturbation of the control from $u(t)$ to $u(t)+\delta u(t)$.
By keeping only the first order terms, we obtain the equation satisfied by $\delta\psi$:
\begin{eqnarray*}
\frac{d}{dt}\delta\psi=-i\left(g(\omega)H_0+u(t)f(\theta)H_1\right)\delta\psi\\
\ \ \  -i\delta u(t)f(\theta)H_1\vert\psi_{\omega,\theta}(t)\rangle,\ \ \ \ \ \ \ \\
\delta\psi(0)=0. \ \ \ \ \ \ \ \  \ \ \ \ \ \ \ \ \ \ \ \ \ \ \ \ \ \ \ \ \ \ \ \
\end{eqnarray*} Let $U_{\omega,\theta}(t)$ be the propagator corresponding to \eqref{app-eq:sch}. Then, $U_{\omega,\theta}(t)$ satisfies
$$\frac{d}{dt}U_{\omega,\theta}(t)=-iH_{\omega,\theta}(t)U_{\omega,\theta}(t),\quad U(0)=\textrm{Id}.$$
Therefore,
\begin{eqnarray}
\delta\psi(T)=-iU_{\omega,\theta}(T)\int_0^T\delta u(t)U_{\omega,\theta}^\dagger(t)f(\theta)H_1\vert\psi_{\omega,\theta}(t)\rangle dt\nonumber \\
\ =-iU_{\omega,\theta}(T)\int_0^TU_{\omega,\theta}^\dagger(t)f(\theta)H_1U_{\omega,\theta}(t) \delta u(t)dt~ \vert\psi_0\rangle.\label{app-eq:deltapsi}
\end{eqnarray}
Using \eqref{app-eq:deltapsi}, we compute $J_{\omega,\theta}(u+\delta u)$ as follows
\begin{eqnarray}
&J_{\omega,\theta}(u+\delta u)-J_{\omega,\theta}(u)\nonumber\ \ \ \ \ \  \  \ \ \ \ \ \ \ \ \ \ \ \ \ \ \ \ \ \ \ \ \ \ \ \\
\approx&2\Re\left(\langle\psi_{\omega,\theta}(T)\vert\psi_{\textrm{target}}\rangle\langle\psi_{\textrm{target}}\vert\delta\psi(T)\right)\nonumber \ \ \ \ \  \ \ \  \ \ \  \ \ \ \ \ \ \ \ \\
=&2\Re\left(-i\langle\psi_{\omega,\theta}(T)\vert\psi_{\textrm{target}}\rangle\langle\psi_{\textrm{target}}\vert \int_0^TV(t) \delta u(t)dt~ \vert\psi_0\rangle\right)\nonumber\\
=&\int_0^T2\Im\left(\langle\psi_{\omega,\theta}(T)\vert\psi_{\textrm{target}}\rangle\langle\psi_{\textrm{target}}\vert V(t)\vert\psi_0\rangle\right)\delta u(t)dt,  \ \ \ \label{app-eq:dJ}
\end{eqnarray}where $\Re(\cdot)$ and $\Im(\cdot)$ denote, respectively, the real and imaginary parts of a complex number, and $V(t)=U_{\omega,\theta}(T)U_{\omega,\theta}^\dagger(t)f(\theta)H_1U_{\omega,\theta}(t)$.

Recall also that the definition of the gradient implies that
\begin{eqnarray}
&J_{\omega,\theta}(u+\delta u)-J_{\omega,\theta}(u)\nonumber\ \ \ \ \ \ \ \ \ \ \  \  \\
=&\langle \nabla J_{\omega,\theta}(u),\delta u\rangle_{L^2([0,T])}+o(\Vert\delta u\Vert)\nonumber\\
=&\int_0^T \nabla J_{\omega,\theta}(u)\delta u(t)dt+o(\Vert\delta u\Vert).\label{app-eq:dJ2}
\end{eqnarray}Therefore, by identifying \eqref{app-eq:dJ} with \eqref{app-eq:dJ2}, we obtain
\begin{equation}\label{app-eq:gradJ}
 \nabla J_{\omega,\theta}(u)=2\Im\left(\langle\psi_{\omega,\theta}(T)\vert\psi_{\textrm{target}}\rangle\langle\psi_{\textrm{target}}\vert V(t)\vert\psi_0\rangle\right).
\end{equation}
The
gradient flow method can be generalized to the case with $M>1$ as
shown in \emph{Algorithm 1}.

\begin{algorithm}
\caption{Gradient flow based iterative learning}
\label{ModifiedGradientFlow}

\begin{algorithmic}[1]

\State Set the index of iterations $k=0$

\State Choose a set of arbitrary controls $u^{k=0}=\{u_{m}^{0}(t),\
m=1,2,\ldots,M\}, t \in [0,T]$

\Repeat {\ (for each iterative process)}

\Repeat {\ (for each training samples $n=1,2,\ldots,N$)}

\State Compute the propagator $U_{n}^{k}(t)$ with the control
strategy $u^{k}(t)$


\Until {\ $n=N$}

\Repeat {\ (for each control $u_{m}(m=1,2,\ldots,M)$ of the control
vector $u$)}

\State
$\delta_m^{k}(t)=2\Im\left(\langle\psi_{\omega_n,\theta_n}(T)\vert\rho_{\textrm{target}} V_{\omega_n,\theta_n}(t)\vert\psi_0\rangle\right)$ where $V_{\omega_n,\theta_n}(t)=U_{\omega_n,\theta_n}(T)U_{\omega_n,\theta_n}^\dagger(t)f(\theta_n)H_mU_{\omega_n,\theta_n}(t)$ and $\rho_{\textrm{target}}=\vert\psi_{\textrm{target}}\rangle\langle\psi_{\textrm{target}}\vert$

\State $u_{m}^{k+1}(t)=u_{m}^{k}(t)+\eta^{k} \delta_{m}^{k}(t)$

\Until {\ $m=M$}

\State $k=k+1$

\Until {\ the learning process ends}

\State The optimal control strategy
$u^{*}=\{u_{m}^*\}=\{u_{m}^{k}\}, \ m=1,2,\ldots,M$

\end{algorithmic}
\end{algorithm}

\begin{remark}
The numerical solution of the control design using \emph{Algorithm
1} is always difficult with a time varying continuous control
strategy $u(t)$. In a practical implementation, we usually
divide the time interval $[0,T]$ equally into a number of time
slices $\triangle t$ and assume that the controls are constant
within each time slice. Instead of $t \in [0,T]$ the time
index will be $t_{q}=qT/Q$, where $Q=T/\triangle t$ and
$q=1,2,\ldots,Q$.
\end{remark}

\section{SLC for three-level quantum systems with uncertainties}\label{Sec4}
In this section, we demonstrate the application of the
proposed SLC method to a $V$-type three-level quantum systems with Hamiltonian uncertainties.

\subsection{Control of a $V$-type quantum system}
We consider a $V$-type quantum system and
demonstrate the SLC design process. Assume that the initial state is
$|\psi(t)\rangle=c_{1}(t)|1\rangle+c_{2}(t)|2\rangle+c_{3}(t)|3\rangle$.
Let $C(t)=(c_{1}(t),c_{2}(t),c_{3}(t))$ where the $c_i(t)$'s are complex numbers. We have
\begin{equation}
i\dot{C}(t)=(g(\omega(t))H_{0}+f(\theta(t))H_{u}(t))C(t).
\end{equation}
We take $H_{0}=\textrm{diag}(1.5, 1, 0)$ and choose $H_{1}$, $H_{2}$,
$H_{3}$ and $H_{4}$ as follows \cite{Hou et al 2012}:
\begin{equation}\label{h0}
H_{1}=
\left(%
\begin{array}{ccc}
  0 & 1 & 0 \\
  1 & 0  & 0 \\
  0 & 0  & 0 \\
\end{array}%
\right), \ H_{2}=
\left(%
\begin{array}{ccc}
  0 & -i & 0 \\
  i & 0  & 0 \\
  0 & 0  & 0 \\
\end{array}%
\right), \nonumber \end{equation}

\begin{equation}
\ H_{3}=
\left(%
\begin{array}{ccc}
  0 & 0 & 1 \\
  0 & 0  & 0 \\
  1 & 0  & 0 \\
\end{array}%
\right), \ \ \ H_{4}=
\left(%
\begin{array}{ccc}
  0 & 0 & -i \\
  0 & 0  & 0 \\
  i & 0  & 0 \\
\end{array}%
\right).
\end{equation}
%
After we sample the uncertainties, every sample can be described as follows:
\begin{equation}\label{general3level}
\left(%
\begin{array}{c}
  \dot{c_{1}}(t) \\
  \dot{c_{2}}(t) \\
  \dot{c_{3}}(t) \\
\end{array}%
\right)=
\left(%
\begin{array}{ccc}
  -1.5g(\omega) i & F_{1}(\theta)  & F_{2}(\theta) \\
  F^{*}_{1}(\theta) & -g(\omega) i  & 0 \\
  F^{*}_{2}(\theta) & 0 & 0 \\
\end{array}%
\right) \left(%
\begin{array}{c}
  c_{1}(t) \\
  c_{2}(t) \\
  c_{3}(t) \\
\end{array}%
\right)
\end{equation}
where $F_{1}(\theta)=f(\theta)[u_{2}(t)-iu_{1}(t)]$, $F_{2}(\theta)=f(\theta)[u_{4}(t)-iu_{3}(t)]$, $\omega\in [-\Omega, \Omega]$ and $\theta \in [-\Theta,
\Theta]$. $\Omega \in [0,1]$ and $\Theta \in [0,1]$ are given
constants.

To construct an augmented system for the training step of the
SLC design, we choose $N$ training samples
(denoted as $n=1, 2, \ldots, N$) through sampling the uncertainties as follows:
\begin{equation}\label{3level-element1}
\left(%
\begin{array}{c}
  \dot{c}_{1,n}(t) \\
  \dot{c}_{2,n}(t) \\
  \dot{c}_{3,n}(t) \\
\end{array}%
\right)=B_{n}(t)\left(%
\begin{array}{c}
  c_{1,n}(t) \\
  c_{2,n}(t) \\
  c_{3,n}(t) \\
\end{array}%
\right),\end{equation}
\begin{equation}B_{n}(t)=
\left(%
\begin{array}{ccc}
  -1.5g(\omega_{n}) i & F_{1}(\theta_{n}) & F_{2}(\theta_{n}) \\
  F^{*}_{1}(\theta_{n}) & -g(\omega_{n}) i  & 0 \\
  F^{*}_{2}(\theta_{n}) & 0 & 0 \\
\end{array}%
\right)\nonumber,
\end{equation}
where $F_{1}(\theta_{n})=f(\theta_{n})[u_{2}(t)-iu_{1}(t)]$, $F_{2}(\theta_{n})=f(\theta_{n})[u_{4}(t)-iu_{3}(t)]$. We assume that $\omega_{n} \in [-\Omega, \Omega]$ and $\theta_{n} \in
[-\Theta, \Theta]$ have uniform distributions. Now the
objective is to find a robust control strategy $u(t)=\{u_{m}(t), m=1,2,
3,4\}$ to drive the quantum system from
$|\psi_{0}\rangle=\frac{1}{\sqrt{3}}(|1\rangle+|2\rangle+|3\rangle)$
(i.e.,
$C_{0}=(\frac{1}{\sqrt{3}},\frac{1}{\sqrt{3}},\frac{1}{\sqrt{3}})$)
to $|\psi_{\text{target}}\rangle=|3\rangle$ (i.e., $C_{\text{target}}=(0,0,1)$).

If write (\ref{3level-element1}) as
$\dot{C}_{n}(t)=B_{n}(t)C_{n}(t)$ ($n=1,2,\ldots, N$), we can
construct the following augmented equation
\begin{equation}\label{augmented-equation-5samples3level}
\left(%
\begin{array}{c}
  \dot{C}_{1}(t) \\
  \dot{C}_{2}(t) \\
  \vdots \\
  \dot{C}_{N}(t) \\
\end{array}%
\right)=
\left(%
\begin{array}{cccc}
  B_{1}(t) & 0 & \cdots & 0  \\
  0 & B_{2}(t) & \cdots & 0  \\
  \vdots & \vdots & \ddots & \vdots  \\
  0 & 0 & \cdots & B_{N}(t)  \\
\end{array}%
\right) \left(%
\begin{array}{c}
  C_{1}(t) \\
  C_{2} (t)\\
  \vdots \\
  C_{N} (t)\\
\end{array}%
\right).
\end{equation}
For this augmented equation, we use the training step to learn
an optimal control strategy $u(t)$ to maximize the following
performance function
\begin{equation}
J(u)=\frac{1}{N}\sum_{n=1}^{N}\vert \langle
C_{n}(T)|C_{\text{target}}\rangle\vert^{2}.
\end{equation}

Now we employ \emph{Algorithm 1} to find the optimal control strategy
$u^{*}(t)=\{u^{*}_{m}(t), m=1,2,3,4\}$ for this augmented
system. Then we apply the optimal control strategy to other
samples to evaluate its performance.

\subsection{Numerical example}
For the numerical experiments on a $V$-type quantum system \cite{You and Nori 2011},
we use the parameter settings listed as follows: the initial state
$C_{0}=(\frac{1}{\sqrt{3}},\frac{1}{\sqrt{3}},\frac{1}{\sqrt{3}})$,
and the target state $C_{\text{target}}=(0,0,1)$; The end time is $T=5$
and the total time interval $[0,T]$ is equally discretized into
$Q=200$ time slices with each time slice $\Delta
t=(t_{q}-t_{q-1})|_{q=1,2,\ldots,Q}=T/Q=0.025$; The learning rate is
$\eta^{k}=0.2$; The control
strategy is initialized with $u^{k=0}(t)=\{u^{0}_{m}(t)=\sin t,
m=1,2,3,4 \}$.

First, we assume that there exist only uncertainty $g(\omega(t))$ (i.e., $f(\theta(t))\equiv 1$), $g(\omega(t))=1-\omega \cos t$, $\Omega=0.28$ and $\omega$ has a uniform
distribution in the interval $[-0.28, 0.28]$. To construct an augmented system for the
training step, we have the training samples for
this $V$-type quantum system as follows
\begin{equation}
\left\{ \begin{split}
& g(\omega_{n})=1-0.28+\frac{0.28(2n-1)}{7},\\
& f(\theta_{n})=1, \\
\end{split}\right.
\end{equation}
where $n=1,2,\ldots,7$. The training
performance for this augmented system is shown in Fig. 1. It is
clear that the
learning process converges to a quite satisfying stage with only
about $300$ iterations. The optimal control strategy is
demonstrated in Fig. 2, which is compared with the initial one.
To test the optimal control strategy obtained from the training
step using the augmented system, we randomly choose $200$
samples through sampling the uncertainty $g(\omega(t))$ and demonstrate the control performance
in Fig. 3. The average fidelity is 0.9989.

\begin{figure}\label{fig1}
\centering
\includegraphics[width=3.6in]{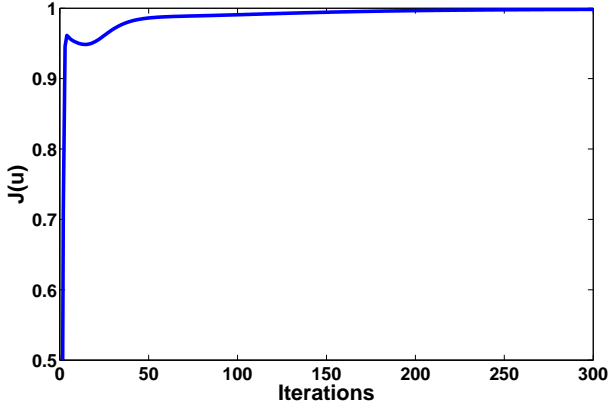}
\caption{Training performance to find the optimal control
strategy by maximizing $J(u)$ for the $V$-type quantum system with
only uncertainty $g(\omega(t))$ where $\omega(t) \in [-0.28, 0.28]$.}
\end{figure}

\begin{figure}\label{fig2}
\centering
\includegraphics[width=3.75in]{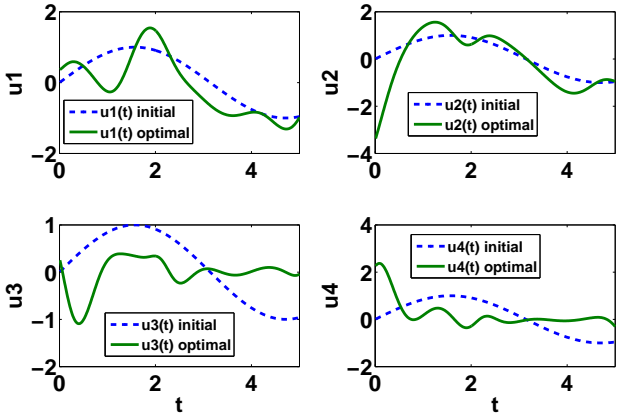}
\caption{The learned optimal control strategy with maximized J(u)
for the $V$-type quantum system with
only uncertainty $g(\omega(t))$ where $\omega(t) \in [-0.28, 0.28]$.}
\end{figure}

\begin{figure*}\label{fig3}
\centering
\includegraphics[width=5.08in]{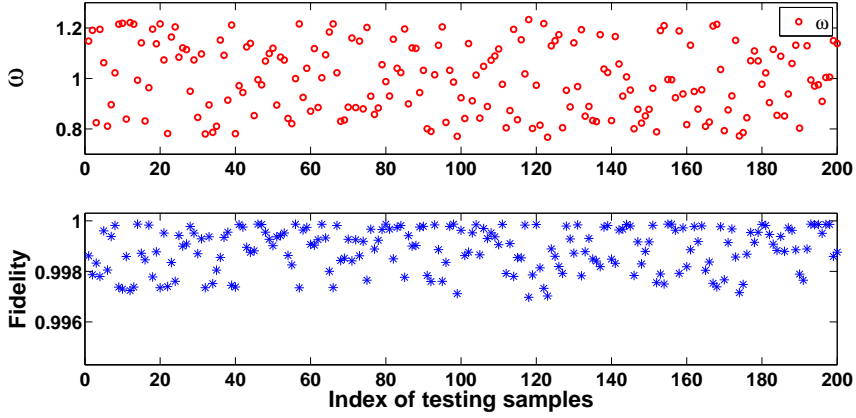}
\caption{The testing performance (with respect to fidelity) of the
learned optimal control strategy for the $V$-type quantum system with
only uncertainty $g(\omega(t))$ where $\omega(t) \in [-0.28, 0.28]$. For the 200 testing samples, the mean fidelity is 0.9989.}
\end{figure*}

Now we consider the more general case that there exist uncertainties  $g(\omega(t))$ and $f(\theta(t))$. Assume
$\Omega=\Theta=0.28$, $g(\omega(t))=1-\omega \cos t$, $f(\theta(t))=1-\theta \cos t$ and both $\omega$ and $\theta$ have uniform
distributions in the interval $[-0.28, 0.28]$.
To construct an augmented system for the
training step, we have the training samples as follows
\begin{equation}
\left\{ \begin{split}
& g(\omega_{n})=1-0.28+\frac{0.28(2\text{fix}(n/7)-1)}{7},\\
& f(\theta_{n})=1-0.28+\frac{0.28(2\text{mod}(n,7)-1)}{7}, \\
\end{split}\right.
\end{equation}
where $n=1,2,\ldots,49$, $\text{fix}(x)=\max \{z\in \mathbb{Z}|z\leq x\}$, $\text{mod}(n,7)=n-7z (z\in \mathbb{Z}\ \text{and}\ \frac{n}{7}-1<z\leq \frac{n}{7} )$ and $\mathbb{Z}$ is the set of integers.
The training
performance for this augmented system is shown in Fig. 4. The optimal control strategy is
presented in Fig. 5.
To test the optimal control strategy obtained from the training
step using the augmented system, we randomly choose $200$
samples through sampling the uncertainties $g(\omega(t))$ and $f(\theta(t))$ whose control performance is presented
in Fig. 6. The average fidelity is 0.9901.
These numerical results show that the proposed SLC method
using an augmented system for training is effective for control
design of quantum systems with Hamiltonian uncertainties.

\begin{figure}\label{fig4}
\centering
\includegraphics[width=3.6in]{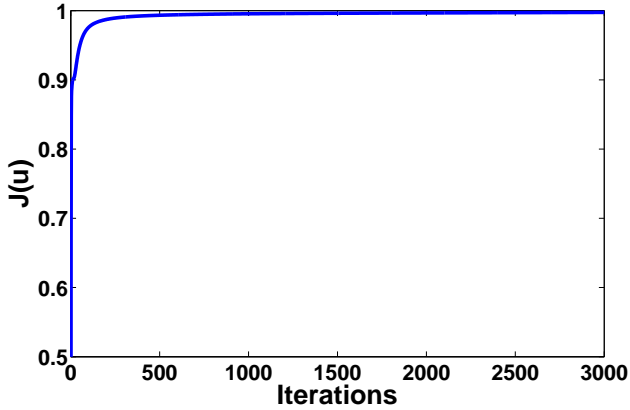}
\caption{Training performance to find the optimal control
strategy by maximizing $J(u)$ for the $V$-type quantum system with
uncertainties $g(\omega(t))$ and $f(\theta(t))$ where $\omega(t) \in [-0.28, 0.28]$ and $\theta(t) \in [-0.28, 0.28]$.}
\end{figure}

\begin{figure}\label{fig5}
\centering
\includegraphics[width=3.75in]{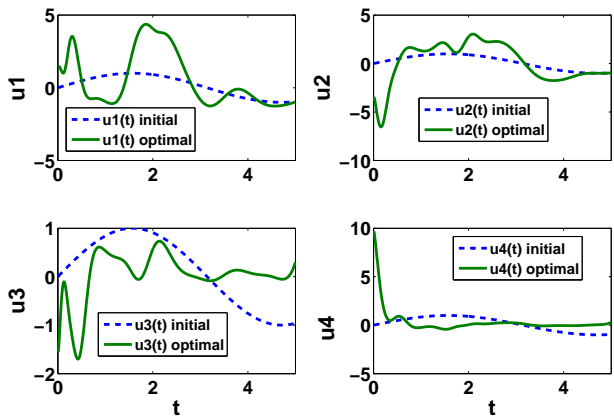}
\caption{The learned optimal control strategy with maximized J(u)
for the $V$-type  quantum system with
uncertainties $g(\omega(t))$ and $f(\theta(t))$ where $\omega(t) \in [-0.28, 0.28]$ and $\theta(t) \in [-0.28, 0.28]$.}
\end{figure}

\begin{figure*}\label{fig6}
\centering
\includegraphics[width=5.68in]{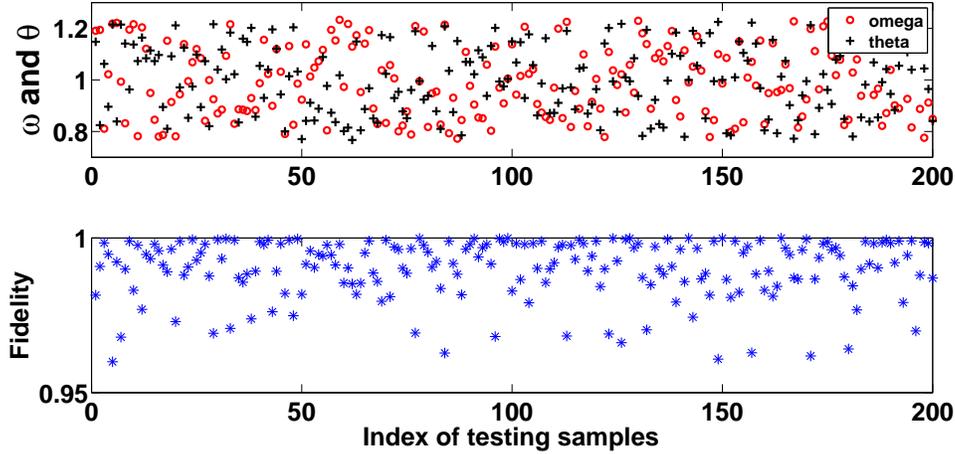}
\caption{The testing performance (with respect to fidelity) of the
learned optimal control strategy for the $V$-type  quantum system with
uncertainties $g(\omega(t))$ and $f(\theta(t))$ where $\omega(t) \in [-0.28, 0.28]$ and $\theta(t) \in [-0.28, 0.28]$.
For the 200 testing samples, the mean fidelity is 0.9901.}
\end{figure*}

\section{Conclusion}\label{Sec5}
In this paper, we presented a systematic numerical methodology for control
design of quantum systems with Hamiltonian uncertainties. The proposed sampling-based learning control method
includes two steps of ``training" and ``testing and evaluation".
In the training step, the control is learned using a gradient flow based
learning and optimization algorithm for an augmented system
constructed from samples. In the process of testing and evaluation, the control
obtained in the first step is evaluated for additional
samples. The results show the effectiveness of the SLC method for
control design of quantum systems with Hamiltonian uncertainties.

\section*{Acknowledgment}
The authors would like to thank Prof. Herschel Rabitz
for his helpful discussion.

\end{document}